\documentclass[%
 aip,
rsi,%
amsmath,amssymb,
preprint,%
]{revtex4-1}

\usepackage{graphicx}
\usepackage{dcolumn}
\usepackage{bm}

\begin{document}


\title{On the miscibility gap of Cu-Ni system}

\author{Yusuke Iguchi}
\email[Corresponding author: ]{iguchi.yusuke@atomki.mta.hu}
\altaffiliation[Present Affiliation: ]{Institute for Nuclear Research, 
 Hungarian Academy of Sciences, Bem ter 18/c., Debrecen, H-4026 HUNGARY}

 \author{G\'abor Katona}
 \author{Csaba Cserh\'ati}
 \author{G\'abor  Langer}
    
\author{Zolt\'an Erd\'elyi}%
\homepage{http://web.unideb.hu/zerdelyi}

\affiliation{ Department of Solid State Physics, University of
  Debrecen, Bem ter 18/b., Debrecen, H-4026 HUNGARY
}%

\date{\today}

\begin{abstract}
The existence of the miscibility gap in the Cu-Ni system has been
  an issue in both computational and experimental discussions for half
  a century [Chakrabarti \textit{et al., Phase diagrams of binary 
  nickel alloys}, ASM, 1991]. Here we propose a new 
  miscibility gap in the Cu-Ni system measured in nano-layered thin
  films by Secondary Neutral Mass Spectrometry. The maximum of the
  symmetrical miscibility curve is around 800~K at Cu$_{50\%}$Ni$_{50\%}$. 
  To our best knowledge, this is the first experiment determining 
  the miscibility from the measurement of the atomic fraction of
  Copper and Nickel in the whole composition range. Needless to say
  that Ni, Cu and its alloys are important in various fields,
  accordingly this result affects different areas to understand materials sciences.
%
\end{abstract}

                             
                             

\maketitle

A phase diagram is a type of chart used to show conditions -- usually
temperature versus composition in the binary systems \cite{90Mass} --
at which thermodynamically distinct phases can exist at equilibrium. A
miscibility gap is a region in a phase diagram where the mixture of
components exists as two or more phases. The knowledge of phase
diagrams is fundamental for example to construct interaction
potentials for computer simulations and indispensable for material
designing. It is not surprising than that there are still substantial efforts
developing
experimental and theoretical methods of establishing phase relations 
in multicomponent systems.

Cu-Ni alloys are very popular in various fields, e.g.\
corrosion-resistant structural materials, welding, soldering,
resistance and magnetic devices. Cu-Ni is also frequently used as an
ideal system to study the surface segregation and diffusion in
completely miscible alloys. Yet, the phase equilibrium, the
immiscibility in the Cu-Ni system remained a serious and unresolved
issue in the past half a century. In 1957 the miscibility gap, that
closes with critical point at 450~K, was predicted from thermodynamic
analysis of ternary Cu-Ni-Cr system \cite{57Mei}. Later, there were
many reports about measurements of physical properties of Cu-Ni, such
as electrical resistivity, Hall effect \cite{57Kos,61Sch2}, specific
heat \cite{64Gup,65Paw}, M\"{o}ssbauer Effect \cite{71Joh} and Nuclear
Magnetic Resonance \cite{65Bea}. These measurements provided some
\emph{indirect} information about the possible existence of the
miscibility gap. Some structural measurements, such as Transmission
Electron Microscopy \cite{74Mur}, X-ray diffraction
\cite{71Ebe,81Tsa}, Atom Probe Field Ion Microscopy \cite{92Lop} and
Neutron scattering \cite{68Moz,78Vri,80Wag} also suggested a sort of
phase separation, e.g.\ clustering. These experimental studies suggested
critical temperature values. Besides these efforts,
theoreticians created thermodynamic models to predict the Cu-Ni
equilibrium phase diagram. Studies up to 1990s are summarized by
Chakrabarti et al.\ \cite{91Cha}. They listed seven experiments and
nine thermodynamic calculations suggesting critical temperature
values. Five of the seven reported experiments and two of the nine
thermodynamic calculations suggested critical temperature values in
the ``high range'' (see Table 2 in \cite{91Cha}; only six of the seven
experiments are listed in the table as \cite{61Sch2} verified the
results of \cite{57Kos}): between 723 and 923~K. Two of the
experiments and seven of the nine thermodynamic calculations of the
suggested critical temperature values are in the ``low range'' (see
Table 2 in \cite{91Cha}): one 450~K, one 543~K and seven in the range
of 573-667~K. Chakrabarty et al.\ decided to fit seven results
(two experiments, five calculations) which are in the range of
573-667~K by a thermodynamic model, although altogether five
experiments suggested critical temperature values in the high range
and two in the low range. They produced a miscibility curve with a
closing critical temperature of 627~K, which was published in the
``Phase Diagram of Binary Nickel Alloys'' book by Chakrabarti
\cite{91Cha}. This curve was then used in the well-known ``Binary
Alloys Phase Diagrams'' book by Massalski (1990) \cite{90Mass}. From
this time on, their phase diagram has been widely used for the last
decades.

Although probably Chakrabarty's phase diagram is the most widely used
in the literature, researchers has continued to investigate Cu-Ni
interaction due to the importance of the Cu-Ni (based) alloys and
because the published data on thermodynamic mixing functions \cite{91Cha}
demonstrate substantial differences both in the absolute value and
temperature dependence, moreover phase equilibrium in the immiscible
region cannot be considered fully understood. For example in 2007,
Turchanin et al.\ \cite{07Tur} published a thorough summary about the
Cu-Ni system discussing the miscibility gap and its thermodynamic
properties. Recently, Kravets et al.\ \cite{12Kra} wrote a detailed study
about the miscibility gap with the ferromagnetic transition (Nishizawa
horn \cite{79Nis}) in this system. It is worth mentioning that the New
Series of Landolt-B\"ornstein also addresses the question of the Cu-Ni
interaction and states that ``Indirect experimental evidence indicates
the presence of a miscibility gap in the fcc phase at a temperature
somewhere between 450 and 923~K.'' \cite {LB}

Although the published data show substantial differences, there is an
agreement that the miscibility gap cannot be measured directly by
classical metallurgical methods (separation from quenched homogeneous
supersaturated alloys), since the diffusion of the atoms is so slow in 
this temperature range that the equilibrium cannot be achieved in 
reasonable time, moreover, the separation takes place in such a 
microscopic regions that this cannot be observed.

Unlike in previous works, this communication provides
experimental results on the miscibility in the Cu-Ni system obtained
by measuring the atomic fraction of the constituents. To overcome
the problem of slow diffusion, hence long experimental time, we choose the so-called
"diffusion couple technique" to determine the miscibility gap. The idea to use
macroscopic diffusion couples
for constructing phase diagrams was suggested long time ago 
\cite{Kirkaldy}. The method is based on the 
assumption of local equilibrium in the diffusion
zone \cite{Kodentsov}. This implies that each infinitely thin layer of such a
diffusion zone is
in thermodynamic equilibrium with the neighboring layers. This means that the
chemical potential of the different species varies continuously through the product 
layers of the diffusion zone. In the following we suppose that local 
equilibrium is maintained in the diffusion zone.

Accordingly, we used nanolayered materials instead of equilibrating homogeneous
supersaturated alloys: we followed the structural and compositional
changes in thin tri- and bilayers by depth profiling with Secondary
Neutral Mass Spectrometry (SNMS). For details about SNMS, see
Supplementary Material.

In principle, there are two ways to move a binary phase separating
system across the solubility curve: i) keeping the temperature but
changing the average composition; ii) changing the temperature but
keeping the average composition of the system. Hence several types of
polycrystalline thin film structures were synthesized by sputtering and annealed in
vacuum ($5 \times 10^{-4}$~Pa) at temperatures ranging from 670 to
838~K for various durations: (a)
Ni$_{30nm}$/Cu$_{70nm}$/Ni$_{30nm}$/SiO$_x$, (b)
Ni$_{29nm}$/Cu$_{31nm}$/SiO$_x$, (c) Cu$_{31nm}$/Ni$_{29nm}$/SiO$_x$,
(d) Cu$_{70nm}$/Ni$_{30nm}$/Cu$_{70nm}$/SiO$_x$ and (e)
Cu$_{15nm}$/Ni$_{70nm}$/Cu$_{15nm}$/SiO$_x$. Accordingly, the Ni
content of the sample is about 48~at\% for sample (a), 50~at\% for (b)
and (c), moreover 18~at\% for (d) and 72~at\% for (e),
respectively. The samples were analyzed by an INA-X type SNMS (SPECS
GmbH) equipment with a depth resolution better than 2~nm.


\begin{figure}[t]
\includegraphics[width=8.3cm,clip]{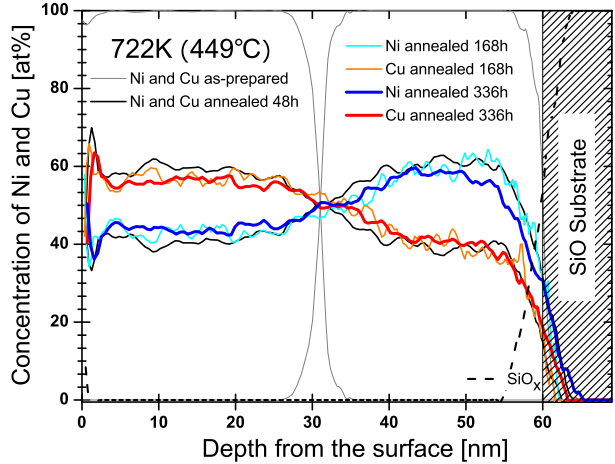}
\caption{Depth profiles recorded by means of SNMS of a sample (c)
Cu$_{31nm}$/Ni$_{29nm}$/SiO$_x$ (Ni average composition 50 at\%;
total film thickness 60 nm)in as-prepared and annealed states (at
722~K for 48, 168, 336 hours). The shaded zone is the SiO$_x$
substrate.}
\label{fig:profile2}
\end{figure}

As an example, Fig.\ \ref{fig:profile2} shows the depth profiles
of a bilayered sample type (c) Cu$_{31nm}$/Ni$_{29nm}$/SiO$_x$ after an
isothermal heat treatment at 722~K for 48, 168 and 336 hours together with the
as-prepared
one. The initially pure Cu/Ni bilayer structure transformed into a Cu-rich/Ni-rich
structure keeping the initial interface positions. This result clearly shows
that $\alpha$1, $\alpha$2 phase separation occurred at 722K with
about 40 at\% ($\alpha$1) and 60 at\% ($\alpha$2) of Nickel. Note that the
equilibrium composition of the different phases which formed due to phase separation
(miscibility gap) in the bilayered samples after annealing for 168h 
at 722K are practically the same: about 42 at\% ($\alpha$1) and 58 at\% ($\alpha$2)
Nickel.

More detailed discussion about equilibrium, interface, grain size after annealing
and technical issues in SNMS can be fined in Supplementary Material.

\begin{figure}[t]
\includegraphics[width=8.3cm,clip]{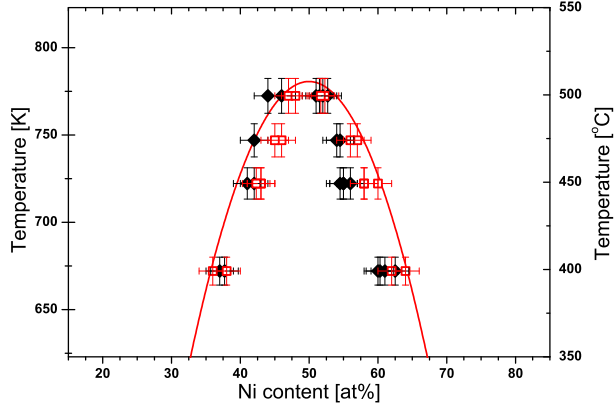}
\caption{ Miscibility gap in Cu-Ni. Closed diamonds are from samples
  type (a) Ni$_{30nm}$/Cu$_{70nm}$/Ni$_{30nm}$/SiO$_x$. Red open
  squares are from samples type (b) Ni$_{29nm}$/Cu$_{31nm}$/SiO$_x$
  and (c)Cu$_{31nm}$/Ni$_{29nm}$/SiO$_x$ (see also text). The miscible
  gap was drawn using simple parabolic curve just to guide the
  eye. (for errors see Supplementary Material) }
\label{fig:phase-diagram}
\end{figure}

Heat treatments of samples type (a)-(c) have been performed at different
temperatures (from 673 to 823~K where is completely paramagnetic region) 
to determine the miscibility gap. 
We also investigated if the free surface (surface segregation) influences the
composition of the Ni- and Cu-rich layers as a function of the depth. In case of the
trilayered samples, type (a), there is one Ni layer on top of the sample
and another one buried under the Cu layer, accordingly we can see the effect of the
free surface on the Ni composition if there is any; in bilayered samples reversing 
the stacking order is needed in order to investigate the same (type (b) and (c)
sample).

In case of the bi- and trilayered samples, we determined the typical duration of
annealing
needed to reach the equilibrium was at least one week. In case of
bilayered samples, we performed heat treatments of 168 to 336 hours (1 to 2 weeks). 
Occasionally, however, much longer heat
treatments were also preformed---e.g.\ 42 days at 722~K---to check whether the system 
really reached the equilibrium.

Figure \ref{fig:phase-diagram} shows the equilibrium composition of
the Ni- and Cu-rich layers of the samples (a), (b) and
(c). Accordingly, at a given temperature a trilayered sample provides
three points: one from the topmost (Ni) layer, one form the middle
(Cu) layer and one from the bottom (Ni) layer. A bilayered sample
delivers two points---one from each layer---, however, its counterpart
with reverse stacking also gives two points, therefore four points
correspond to the bilayer geometry at a given temperature.

\begin{figure}[t]
\includegraphics[width=8.3cm,clip]{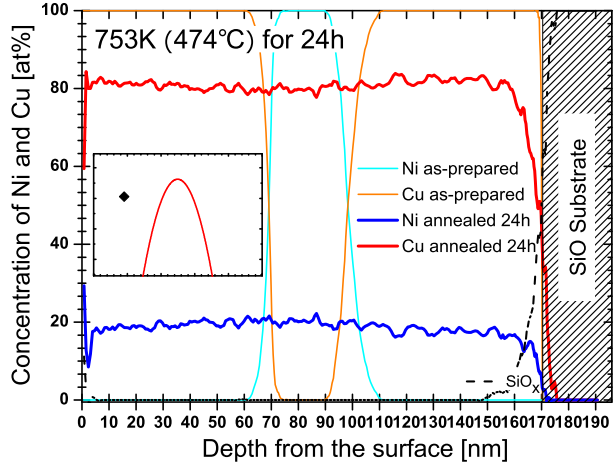}
\caption{Depth profiles recorded by means of SNMS of a sample (d)
  Cu$_{70nm}$/Ni$_{30nm}$/Cu$_{70nm}$/SiO$_x$ (Ni average
  composition 18 at\%; total film thickness 170 nm) in as-prepared
  and annealed states (at 474$^{\circ}$C (753K) for 24 hours). Ni
  contents of this thin film and annealing temperature in this
  experiment are far outside the miscibility gap as shown by a dot
  in the inset. The right shade zone is SiO$_x$ substrate. }
\label{fig:profile3}
\end{figure}

In order to check that the system can reach the homogeneous distribution of
the atoms if the average composition of the sample is outside of the
miscibility gap, we annealed samples type (d) and (e) at 753K just for
24 hours. A typical example of the obtained results is displayed in Fig.\
\ref{fig:profile3}. As can be seen, the
Cu$_{70nm}$/Ni$_{30nm}$/Cu$_{70nm}$/SiO$_x$ sample is almost
homogeneous with sufficiently thicker diffusion length and shorter duration 
than previous results in Fig.1 and 2.


Although it is difficult to be sure that we reached the equilibrium state when using
the "diffusion couple technique", first and last, all of our experimental findings
suggest that what we observed is really
phase separation at or very close to equilibrium cause it has enough volume 
diffusion length $2\sqrt{Dt}$ against that it is sufficientry thinner sample thickness 
with type-A grain boundary diffusion, as an example, diffusion length of Nickel in Copper 
is about 100nm at 722K for 168h\cite{88Neu}.
Ni contents of separated phase are gradually
converging from about 62 and 37 at\% at 672~K to about 50 \% at 800~K
in samples type (a), (b) and (c) as shown in Fig.\
\ref{fig:phase-diagram}. Consequently, the miscibility gap what we
obtained closes at higher temperature than Chakrabarty's one which has
been widely used nowadays. However, the critical temperature we
obtained is similar to the five reported experiments which suggested
critical temperature values in the high range (see Table 2 in
\cite{91Cha}). We repeat here that even Chakrabarty et al.\ themselves listed seven
experiments of which five reported critical temperatures in the "high range" and only 
two in the "low range". Our results support that the closing of the critical
temperature is in the "high range". The striking result in this work is that we could measure the 
whole miscibility gap from \emph{direct} composition measurement in
Cu-Ni system. To our best knowledge, this is the first experiment determining the
miscibility from the measurement of the atomic fraction of Copper and Nickel in the
whole composition range.

Nevertheless, we should not forget about that these measurements were
performed in nano-scaled systems which may rise two points: the possible size
dependence of the miscibility and the validity of the thermodynamic
calculations on the nanoscale. The size dependence of the miscibility
has been predicted in the framework of the Cahn-Hilliard concept
(e.g.\ \cite{09Bur}). However, even Calphad \cite{Calphad} experts
pointed out the problem of the applicability of Calphad type
thermodynamic calculations on the nanoscale. (e.g.\ nano-Calphad
concept \cite{12Kaptay}). According to the nano-Calphad concept ``an
extension of the Calphad method for systems containing at least one
phase (or at least one interface film, complexion) with at least one
of its dimensions being below 100 nm'' is required; without attempting
to be comprehensive: curvature dependence of the interfacial energies,
dependence of interfacial energies on the separation between
interfaces (including the problem of surface melting), role of the
shapes and relative arrangement of phases, role of the substrate (if
such exists), role of segregation. We should not forget about the
possible change of the phase diagram\cite{98Jia, 15Gha, 15Kro} 
with the decreasing size. Concerning these remarks, our samples are just 
touching the challenged dimensions for the determination of the phase diagram.
The total film thickness is 100-170 nm for the tri and 60 nm for the bilayers; 
the individual layer thicknesses are, however, always below 100 nm (from 15 to 70
nm). As we also did some tests with thicker bilayered samples with
total film thickness up to 300 nm which, within the experimental uncertainty, 
resulted the same miscibility values as samples type (a) (total thickness 130 nm)
we think that the results obtained for the trilayered samples give an equilibrium
miscibility gap which is close to the bulk one.

Note that in principle we may also consider that stress modifies the
miscibility gap. Several groups investigated this in both bulk and
thin film samples \cite{OhtaniJPhaseEq2001,DeibukSPQEO2002} and the
elastic energy effect on the spinodal decomposition
\cite{CahnActaMet1963, 78Jan}. The simple conclusion is that stress decreases
the melting point and the critical temperature where the miscibility
curve closes if there is no ordering phase (we did not observe
ordering in our Cu-Ni thin films). According to these studies, however,
the stress does not influence the phase diagram significantly.
 
Stress in our sample may originate mainly from e.g.\ lattice
mismatch, difference in thermal expansions in bimetallic strips and
net volume flux. 1) Lattice mismatch can be neglected because our
samples are polycrystalline (not epitaxial) and there is very small lattice 
mismatches during the heating CuNi compounds. 2) Bimetallic strip effect
between the film and substrate can also be negligible as Tanaka et.\ al
\cite{TanakaJSocMatSci2004} demonstrated that there was nearly zero
stress in in-situ strain measurement of Cu thin film on the SiO
substrate at 300 to 500$^\circ$C during the heating. This is so because Young's
modulus
and linear thermal expansion coefficient of Cu and Ni are very
close. 3) Atomic volumes of Cu and Ni are also very close, so net
volume flux should not be very important; even if it is at the
beginning of the intermixing, in equilibrium the stress should be
relaxed. Summarizing these points above the miscibility gap we measured rather
corresponds
to the stress-free case. 


Eventual conclusion of this work is that clear miscibility gap of
Cu-Ni system with direct measurement of composition was determined on
bi- and tri-layered thin films by means of SNMS. The measured curve
closes around 800~K with almost symmetrical shape. 

We also presented a method how to determine a phase diagram with nano-layered 
thin films using SNMS technique. It was also proved qualitatively that the 
miscibility gap we propose should be the same as the bulk one 
because there is neither obvious size nor stress effect.

These results will presumably set off
huge discussions with impact not only on the Cu-Ni system but also
other binary, ternary system, for example in material designing, nano
materials science, phase simulation, etc.

\section*{SUPPLEMENTARY MATERIAL}
See supplementary material for the detailed discussion about equilibrium, interface, 
structures, grain size after annealing and technical issues.

\begin{acknowledgments}
This work was supported in part by the OTKA Board of Hungary (No.\
NF101329) and by TAMOP 4.2.2.A-11/1/KONV- 2012-0036 project
(implemented through the New Hungary Development Plan co-financed by
the European Social Fund, and the European Regional Development Fund).
Y. I. acknowledge Hungarian 
Academy of Sciences Postdoctoral Fellowship Programme.
\end{acknowledgments}

\nocite{*}
\bibliography{CuNiAPLmain.bib}

\end{document}